\documentclass[twocolumn,floatfix,superscriptaddress,a4paper,showpacs,showkeys,nofootinbib,reprint,prc,noeprint]{revtex4-1}
\usepackage{epsfig}
\usepackage{latexsym}
\usepackage[colorlinks=true,linktocpage=true,linkcolor=blue,citecolor=blue,allcolors=blue]{hyperref}
\usepackage{url}
\usepackage[utf8]{inputenc}
\usepackage{enumerate}
\usepackage{color}
\usepackage{xcolor}

\usepackage{xfrac}

\usepackage{amsmath}
\usepackage{amssymb}

\usepackage[english]{babel}
\usepackage{hyperref}
\usepackage{url}
\usepackage{needspace}
\usepackage{float}
\hypersetup{
   colorlinks=true, 
  linktoc=all,     
  linkcolor=blue,  
}

\begin{document}

 \title{Explanation of the observed violation of isospin symmetry \\ in relativistic nucleus-nucleus reactions}

\author{Tom Reichert}
\affiliation{Institut f\"{u}r Theoretische Physik, Goethe-Universit\"{a}t Frankfurt, Max-von-Laue-Str. 1, D-60438 Frankfurt am Main, Germany}
\affiliation{Frankfurt Institute for Advanced Studies, Ruth-Moufang-Str. 1,  60438 Frankfurt am Main, Germany}
\affiliation{Helmholtz Research Academy Hesse for FAIR (HFHF), GSI Helmholtzzentrum f\"ur Schwerionenforschung GmbH, Campus Frankfurt, Max-von-Laue-Str. 12, 60438 Frankfurt am Main, Germany}

\author{Jan Steinheimer}
\affiliation{GSI Helmholtzzentrum f\"ur Schwerionenforschung GmbH, Planckstr. 1, D-64291 Darmstadt, Germany}
\affiliation{Frankfurt Institute for Advanced Studies, Ruth-Moufang-Str. 1,  60438 Frankfurt am Main, Germany}

\author{Marcus Bleicher}
\affiliation{Institut f\"{u}r Theoretische Physik, Goethe-Universit\"{a}t Frankfurt, Max-von-Laue-Str. 1, D-60438 Frankfurt am Main, Germany}
\affiliation{Helmholtz Research Academy Hesse for FAIR (HFHF), GSI Helmholtzzentrum f\"ur Schwerionenforschung GmbH, Campus Frankfurt, Max-von-Laue-Str. 12, 60438 Frankfurt am Main, Germany}

\date{\today}

\begin{abstract}
The violation of isospin symmetry in nucleus-nucleus reactions, as shown in the ratio ${R_K=(K^++K^-)/(K^0+\bar{K}^0)}$ presented by NA61/SHINE, can be understood by introducing results from color-string fragmentation in $e^+e^-$ to nuclear reactions. This novel input allows for a consistent description of the $e^+e^-$ data, proton+proton data and finally nucleus-nucleus data at all investigated energies. We conclude that the observed isospin violation in nucleus-nucleus reactions is explained by asymmetric production of up- and down-quarks in the elementary color field fragmentation process.
\end{abstract}

\maketitle

\section{Introduction}
Recently, the NA61/SHINE experiment has reported data collected by various experiments on the ratio ${R_K=(K^+ + K^-)/(K^0 + \bar K^0)}$ in nucleus-nucleus reactions over a broad range in energies \cite{NA61SHINE:2023azp,Brylinski:2024uei}. The surprising result was that the ratio $R_K=1.144 \pm 0.026>1$ (averaged over all available energies) is larger than one, indicating that more charged (anti-)Kaons are produced than neutral (anti-)Kaons. This violates the expectation for isospin symmetric nuclear matter\footnote{The quark contents of the Kaons are: $K^+ = (u\bar s)$, $K^0 = (d\bar s)$, $K^- = (\bar u s)$, $\bar K^0 = (\bar d s)$. Thus, one would assume to find more neutral (anti-)Kaons due to the initial $d>u$ quark asymmetry induced by the initial neutron ($udd$) $>$ proton ($uud$) asymmetry.}  which makes up the original colliding atomic nuclei which leads to $R_K\leq 1$. Following these arguments, the NA61/SHINE collaboration stated the observation of ``Evidence of isospin-symmetry violation in high-energy collisions of atomic nuclei'' \cite{NA61SHINE:2023azp}.

A first theoretical exploration of the origin of this observation was done in \cite{Brylinski:2023nrb,Giacosa:2024bup}. Here, the following effects where explored: Weak interactions leading to an uncertainty in $R_K$ by approx. 0.13 \%, electromagnetic processes were found to be negligible, symmetry breaking due to the differences of the $u$ and $d$ quark masses was studied in the hadron resonance gas model \cite{Vovchenko:2019pjl} in more detail in the case of the $\phi$ decay into charged and neutral Kaons and leads to a small increase in $R_K$ of at most $\approx 3 \%$. It was further speculated that a phase of strongly interacting matter with a significant isospin violation (as suggested in \cite{Pisarski:1983ms}) could exist, or that chiral anomaly effects could lead to an isospin asymmetry \cite{Giacosa:2017pos}. However, up to now, no explanation of the observed asymmetry has been brought forward.

In this letter, we explain the observed isospin asymmetry in nucleus-nucleus reactions and show that it is not related to nuclear effects. Rather, we pin down the origin of the observed asymmetry to the fundamental level of quark production in color fields by relating it to the isospin asymmetry observed in $e^+e^-$ annihilations. 

The general line of arguments that we present, based on observations from elementary $e^+e^-$ reactions, holds qualitatively. To make a quantitative statement we implement the effects into the UrQMD event generator \cite{Bleicher:1999xi} and compare $e^+e^-$, proton+proton and nucleus-nucleus reactions to the available experimental data. 

\section{Electron-positron annihilations}
Electron-positron annihilations allow to explore the details of the color field fragmentation to hadrons in great detail. This is relevant for nuclear reactions, because the initial stages of the reaction and also high mass objects are described by the formation and fragmentation of color fields with the same properties as in $e^+e^-$ but with different color charge realizations at the string ends (namely quarks as $[3]$ and diquarks as $[\bar 3]$ as compared to quarks and anti-quarks in $e^+e^-$). The fragmentation of the color field itself is identical to $e^+e^-$. Recently, the BESIII collaboration has published data \cite{BESIII:2022zit,BESIII:2025mbc} on $e^+e^-\rightarrow K^\pm + X$ and $e^+e^-\rightarrow K^0_S + X$ at an energy around 3 GeV. The string mass range explored by BESIII is exactly in the region of the color string masses excited in nuclear collisions and allows therefore to pin down its parameters with high precision. The production process of the Kaons is then given by $e^+e^-\rightarrow q_i\bar q_i \rightarrow\,\mathrm{string}\, \rightarrow K + X$. The production probabilities for the initial $u$, $d$, and $s$ quarks, $q_i$, are given by QED and are therefore proportional to their electric charges $e_i^2$. 
It is important to note that the production of further quark-pairs in the fragmentation of the color field is not a QED process, but is governed by (non-perturbative) QCD. Therefore, different probabilities need to be employed as discussed below.
\begin{figure} [t!]
    \centering
    \includegraphics[width=\columnwidth]{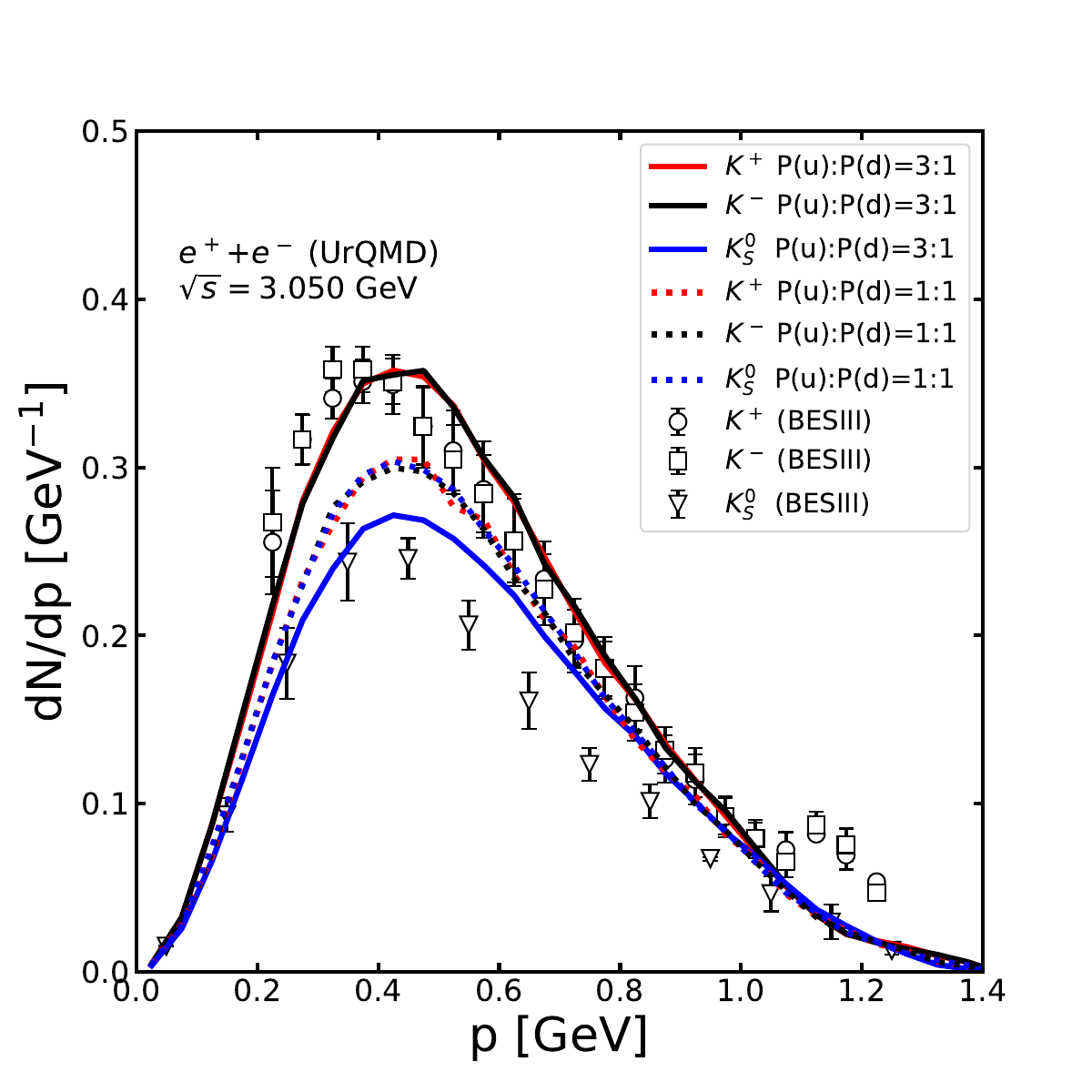}
    \caption{[Color online] Momentum distributions of $K^+$ (red) $K^-$ (black) and $K^0_S=(K^0+\bar K^0)/2$ (blue) in $e^+e^-$ annihilations at $\sqrt s = 3.050$ GeV. The dotted lines show the standard parametrization of the color field fragmentation parameters, the full lines show the refitted parameters which allow for an asymmetry between up- and down-quark production in the color field. The symbols show the experimental data by BESIII \cite{BESIII:2025mbc,BESIII:2022zit}.}
    \label{fig:epem}
\end{figure}
\begin{figure} [t!]
    \centering
    \includegraphics[width=\columnwidth]{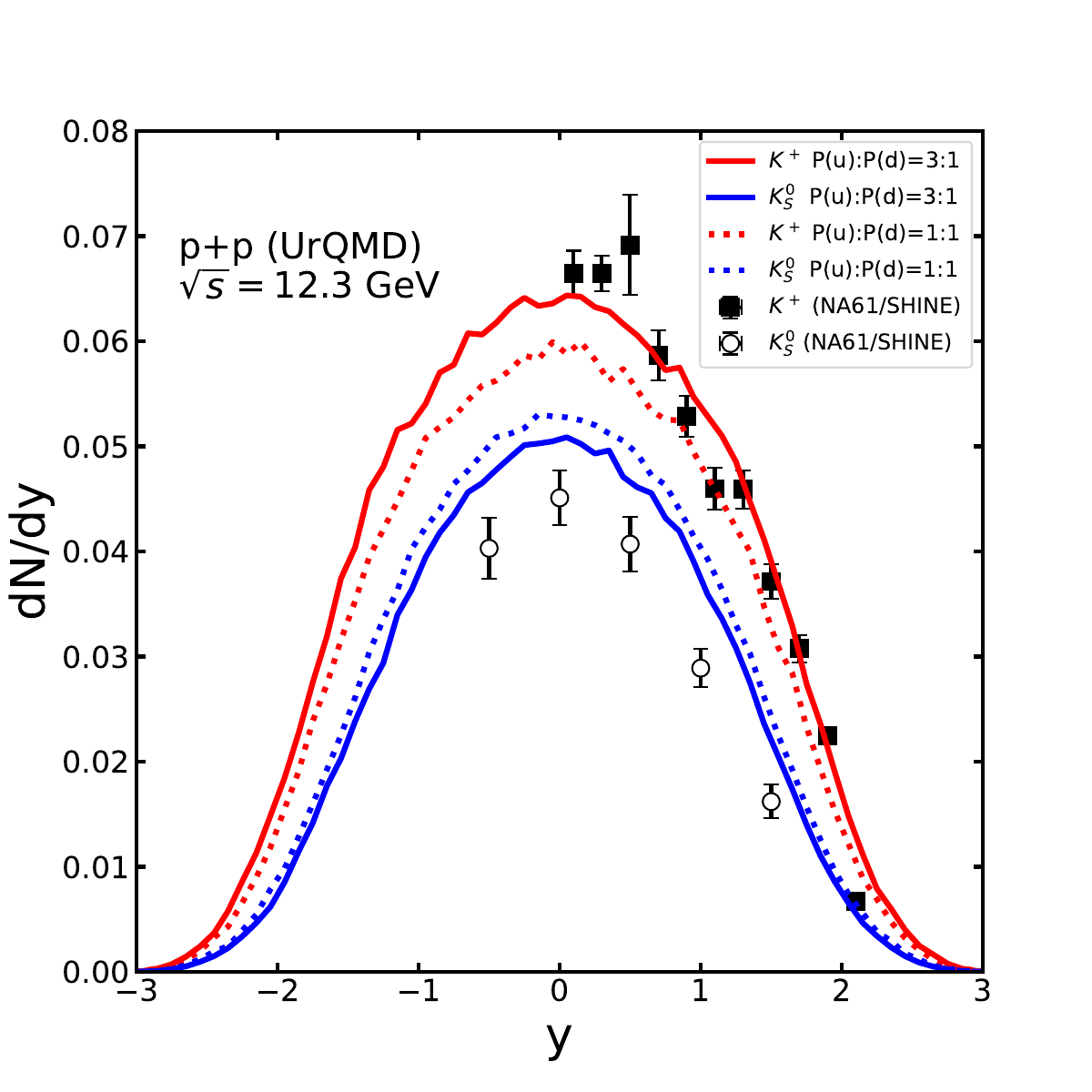}
    \caption{[Color online] Rapidity distributions of $K^+$ (red) and $K^0_S=(K^0+\bar K^0)/2$ (blue) in proton+proton collisions at $\sqrt s = 12.3$ GeV. The dotted lines show the standard parametrization of the color field fragmentation parameters, the full lines show the refitted parameters which allow for an asymmetry between up- and down-quark production in the color field. The symbols show the experimental data by the NA61/SHINE experiment \cite{NA61SHINE:2024eqz,NA61SHINE:2017fne}.}
    \label{fig:pp}
\end{figure}
In Fig. \ref{fig:epem} we show the momentum distributions of $K^+$, $K^-$ and $K^0_S$ in $e^+e^-$ annihilations at $\sqrt s = 3.050$ GeV. The dotted lines show the standard (isospin symmetric) parametrization of the color field fragmentation parameters, the full lines show the refitted parameters which allow for an asymmetry between up- and down-quark production in the color field. The symbols show the experimental data by BESIII \cite{BESIII:2022zit,BESIII:2025mbc}. One clearly observes that the standard parameters assuming same production probabilities, i.e. the same masses, for up- and down-quarks cannot explain the fragmentation process in $e^+e^-$ that shows an asymmetry between charged and neutral Kaons\footnote{The asymmetry in the $R_K$ ratio has been observed at all energies from $\sqrt s =3$ GeV to $\sqrt s =200$ GeV in $e^+e^-$ reactions \cite{ParticleDataGroup:2024cfk}. Here we take the latest results from BESIII as an example.}. However, allowing for a relative increase in the  production probability of up-quarks as compared to down-quarks ($P(u):P(d)\approx 3:1$, instead of the standard assumption $P(u):P(d)=1:1$) results in a very good description of the $e^+e^-$ data. Such an asymmetry in the production probability is plausible due to the mass difference between up- and down-quarks, although the magnitude of the asymmetry is larger than expected from the constituent quark picture. Let us note in passing that also the NNLO fragmentation function approaches cannot describe this asymmetry in the data at low momenta if they assume isospin symmetry \cite{Gao:2024dbv,Gao:2024nkz}. Let us further point out that $P(u):P(d)\approx 3:1$  is also consistent with the observed isospin asymmetry in the charm sector (seen in $e^+e^-$ in the ratio $D^0/D^+ \approx 2.2-2.7$, note that $D^0=(\bar u c)$, $D^+=(\bar d c)$) \cite{ParticleDataGroup:2024cfk}. Thus, one concludes that only an asymmetric production probability of up- and down-quarks in the fragmentation process of the color field is able to describe the measured data in this elementary process.

\section{Kaon production in proton+proton reactions}
Let us now use the $e^+e^-$ adjusted color field fragmentation parameters to explore proton+proton reactions. The only difference between the color field produced in $e^+e^-$ reactions and in proton+proton reactions is the replacement of the color charges at the string ends from $q-\bar q$ to $q-qq$, while the fragmentation parameters of the color field remain the same. 
Fig. \ref{fig:pp} shows the rapidity distributions of $K^+$ and $K^0_S=(K^0+\bar K^0)/2$ in proton+proton collisions at $\sqrt s = 12.3$ GeV. The dotted lines show the standard parametrization of the color field fragmentation parameters, the full lines show the refitted parameters which allow for an asymmetry between up- and down-quark production in the color field. The symbols show the experimental data by the NA61/SHINE experiment \cite{NA61SHINE:2024eqz,NA61SHINE:2017fne}. Again, one finds that the experimentally observed asymmetry between charged Kaons and neutral Kaons in proton+proton reactions can only be described quantitatively with the refitted color field fragmentation parameters including the asymmetric production of up- and down-quarks in the color field. This further supports that the observed isospin asymmetry is encoded at an elementary level.

\begin{figure} [t!]
    \centering
    \includegraphics[width=\columnwidth]{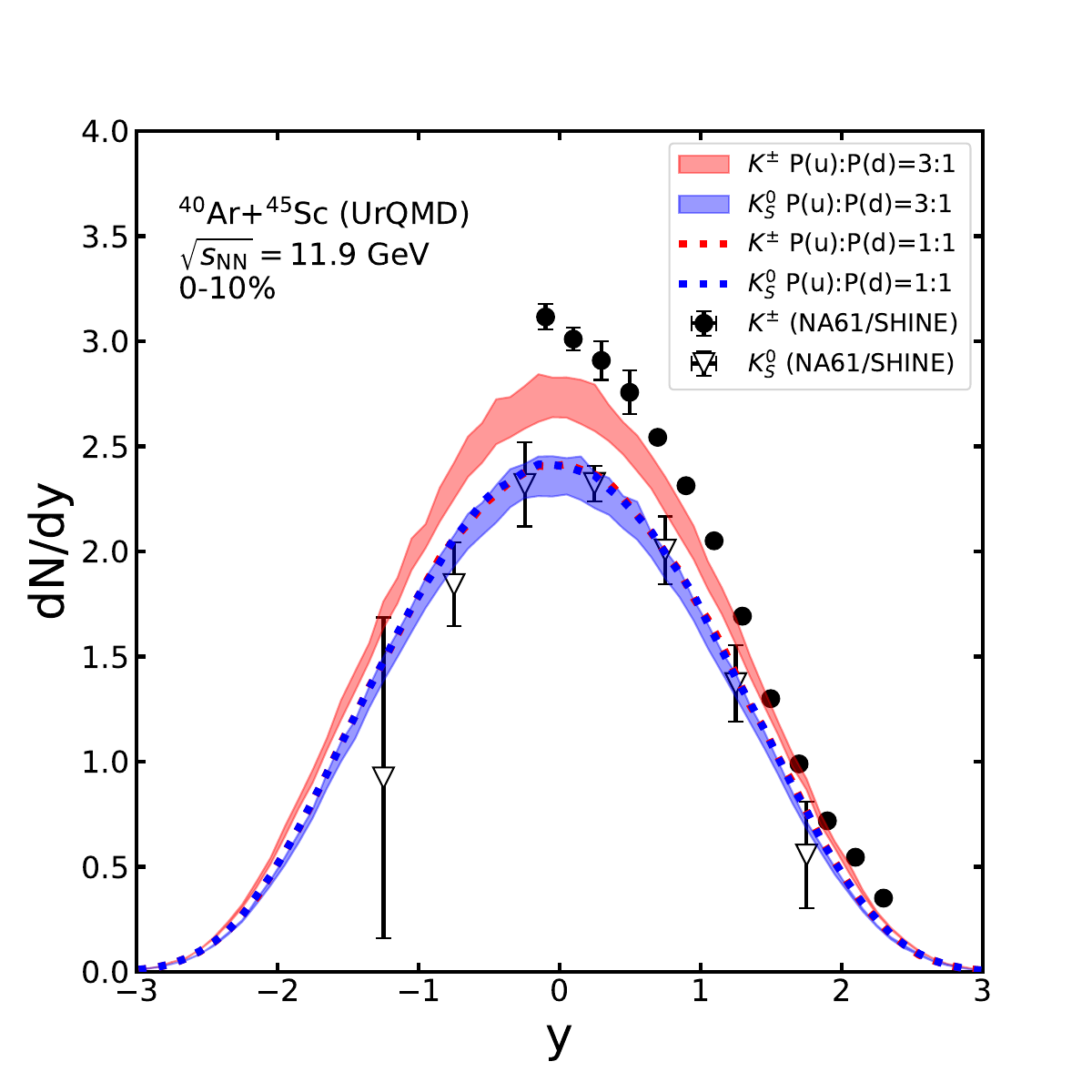}
    \caption{[Color online] Rapidity distributions of $K^\pm = (K^++K^-)/2$ (red) and $K^0_S=(K^0+\bar K^0)/2$ (blue) in 0-10\% central Ar+Sc collisions at $\sqrt{s_\mathrm{NN}} = 11.9$ GeV. The dotted lines show the standard parametrization of the color field fragmentation parameters, the full lines show the refitted parameters which allow for an asymmetry between up- and down-quark production in the color field. The error band in the calculation is due to the estimation of the centrality ($\langle N_\mathrm{wounded}\rangle=60-64$ representing 10\% central collisions). The symbols show the experimental data by the NA61/SHINE experiment \cite{NA61SHINE:2023epu,NA61SHINE:2023azp}. }
    \label{fig:dndy_arsc}
\end{figure}

\begin{figure} [t!]
    \centering
    \includegraphics[width=\columnwidth]{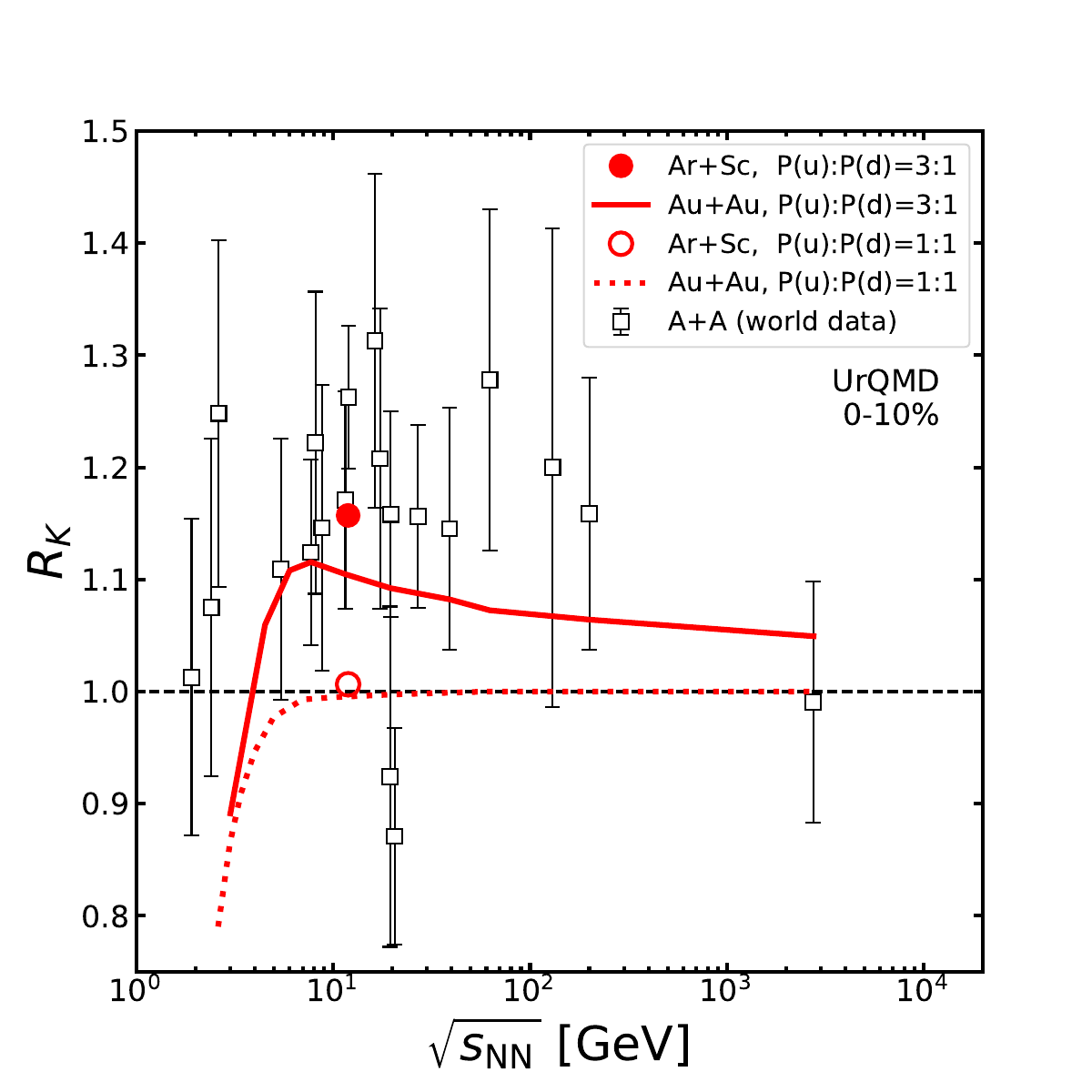}
    \caption{[Color online] Comparison of the refitted simulations with the world data on the ratio $R_K=(K^+ + K^-)/(K^0 + \bar K^0)$. The dotted lines shows central Au+Au reactions with the standard parametrization of the color field fragmentation parameters, the full lines show the refitted parameters which allow for an asymmetry between up- and down quark production in the color field. We further show the results for Ar+Sc collisions (circles) because Ar+Sc has a larger protons/neutron ratio than Au+Au. The black squares show the experimental data in nucleus-nucleus collisions (taken from the compilation in \cite{NA61SHINE:2023epu}). }
    \label{fig:excfct}
\end{figure}
\section{Analysis of the isospin asymmetry in nucleus-nucleus reactions}
Let us finally turn to nucleus-nucleus reactions. In Fig. \ref{fig:dndy_arsc} we show the rapidity distributions of $K^\pm=(K^++ K^-)/2$ and $K^0_S=(K^0+\bar K^0)/2$ in central Ar+Sc collisions at $\sqrt{s_\mathrm{NN}} = 11.9$ GeV. The dotted lines show the standard parametrization of the color field fragmentation parameters, while the full lines show the results using the refitted parameters which allow for an asymmetry between up- and down-quark production in the color field. The error band in the calculation is due to the estimation of the centrality ($\langle N_\mathrm{wounded}\rangle=60-64$ representing 10\% central collisions). The symbols show the experimental data by the NA61/SHINE experiment \cite{NA61SHINE:2023epu,NA61SHINE:2023azp}. This demonstrates that the full extent of the isospin asymmetry in the Ar+Sc reaction can be described based on the refitted color field fragmentation parameters extracted from the $e^+e^-$ reactions.

Fig. \ref{fig:excfct} summarizes our findings by comparing the simulations using the refitted parameters with the world data on the ratio $R_K=(K^+ + K^-)/(K^0 + \bar K^0)$ in nucleus-nucleus reactions (compiled in \cite{NA61SHINE:2023epu}). Also here the inclusion of the asymmetric color field fragmentation parameters for up- and down-quarks results in a good description of the available data.

We conclude that the refitted color field fragmentation parameters allow for a consistent description of the world data on $R_K$. This suggests that the observed isospin violation is not a genuine effect of nucleus-nucleus reactions, but is the result of the isospin asymmetry present in the elementary color field fragmentation.

\section{Conclusion}
The violation of isospin symmetry in nucleus-nucleus reactions presented by NA61/SHINE and observed by a wide range of experiments has been resolved: We have demonstrated that the same isospin violation is already observed in $e^+e^-$ reactions and thus can be traced back to an asymmetry in up- and down-quark production in the color field breakup. Increasing the up-quark production probability, relative to the down-quark, in the color field fragmentation provides consistency with the $e^+e^-$ data on charged and neutral Kaon production. Using this set of adjusted parameters provides a consistent description of proton+proton data. Finally, the measured data on ${R_K=(K^+ + K^-)/(K^0 + \bar K^0)}$ in nucleus-nucleus collisions over a broad range of energies can be described consistently with the same fragmentation parametrization. We conclude that the observed isospin violation in nucleus-nucleus reactions is due to an asymmetric production of up- and down-quarks in the elementary color field fragmentation process. 

Explaining the large asymmetry in the up- and down- quark production in the elementary $e^+e^-$ reactions remains a challenging puzzle to be solved.

\section*{Acknowledgments}
T.R. acknowledges support through the Main-Campus-Doctus fellowship provided by the Stiftung Polytechnische Gesellschaft (SPTG) Frankfurt am Main and moreover thanks the Samson AG for their support.
The computational resources for this project were provided by the Center for Scientific Computing of the GU Frankfurt and the Goethe--HLR.


\end{document}